\documentclass[12pt]{iopart}
\usepackage{graphicx}
\usepackage{units,subfigure,rotate}
\usepackage{color}
\usepackage{longtable}
\usepackage{xspace,textcomp}
\usepackage{amsbsy,amssymb}
\usepackage{dcolumn,bm}							
\newcolumntype{d}[1]{D{.}{.}{#1}} 
\usepackage{hyperref}

\newcommand{\ceossb}{CeOs\texorpdfstring{$_4$}{4}Sb\texorpdfstring{$_{12}$}{12}\xspace}

\begin{document}


\title[Pressure-induced shift of phase boundaries in CeOs\texorpdfstring{$_4$}{4}Sb\texorpdfstring{$_{12}$}{12}]{Pressure-induced shift of effective Ce valence, Fermi energy and phase boundaries in CeOs\texorpdfstring{$_4$}{4}Sb\texorpdfstring{$_{12}$}{12}}

\author{K. G\"{o}tze$^1$\footnote{Present address: Deutsches Elektronen-Synchrotron (DESY), 22607 Hamburg, Germany.}, M.~J. Pearce$^1$\footnote{Present address: Department of Physics, University of Oxford, Clarendon Laboratory, Oxford, OX1 3PU, UK.}, M.~J. Coak$^1$, P.~A. Goddard$^1$, A.~D. Grockowiak$^2$\footnote{Present address: Brazilian Synchrotron Light Laboratory (LNLS), Brazilian Center for Research in Energy and Materials (CNPEM), Campinas, S\~{a}o Paulo, Brazil.}, W.~A. Coniglio$^2$, S.~W. Tozer$^2$, D.~E. Graf$^2$, M.~B. Maple$^3$, P.-C. Ho$^4$, M.~C. Brown$^4$, J. Singleton$^5$}
\address{$^1$ Department of Physics, University of Warwick, Coventry CV4 7AL, UK.}
\address{$^2$ National High Magnetic Field Laboratory, Florida State University, Tallahassee, Florida, 32310, USA.}
\address{$^3$ Department of Physics, University of California, San Diego, La Jolla, CA 92093, USA.}
\address{$^4$ Department of Physics, California State University, Fresno, CA 93740, USA.}
\address{$^5$ National High Magnetic Field Laboratory, Los Alamos National Laboratory, MS-E536, Los Alamos, New Mexico 87545, USA.}
\ead{\mailto{p.goddard@warwick.ac.uk}, \mailto{jsingle@lanl.gov}}

\begin{abstract}

\ceossb, a member of the skutterudite family, has an unusual semimetallic low-temperature $\cal{L}$-phase that inhabits a wedge-like area of the field $H$ - temperature $T$ phase diagram. 
We have conducted measurements of electrical transport and megahertz conductivity on \ceossb single crystals under pressures of up to 3~GPa and in high magnetic fields of up to 41~T to investigate the influence of pressure on the different $H$-$T$ phase boundaries. While the high-temperature valence transition between the metallic $\cal{H}$-phase and the $\cal{L}$-phase is shifted to higher $T$ by pressures of the order of 1~GPa, we observed only a marginal suppression of the $\cal{S}$-phase that is found below 1~K for pressures of up to 1.91~GPa. High-field quantum oscillations have been observed for pressures up to 3.0~GPa and the Fermi surface of the high-field side of the $\cal{H}$-phase is found to show a surprising decrease in size with increasing pressure, implying a change in electronic structure rather than a mere contraction of lattice parameters.
We evaluate the field-dependence of the effective masses for different pressures and also reflect on the sample dependence of some of the properties of \ceossb which appears to be limited to the low-field region.

\end{abstract}

\submitto{\NJP}
\maketitle
\ioptwocol

\section{Introduction}

Filled skutterudites with the chemical formula $RT_4X_{12}$ ($R$=rare-earth element; $T$=Fe, Ru, Os; $X$=P, As, Sb) crystallize in a body-centered cubic structure with a space group $\mathit{Im}\overline{3}$ \cite{jeitschko_1977,braun_1980} and display a large variety of ground states from ferromagnetism in NdOs$_4$Sb$_{12}$ \cite{Ho_2005} to unusual superconductivity in PrOs$_4$Sb$_{12}$ (heavy fermions) \cite{bauer_2002} and LaRu$_4$As$_{12}$ (multiband superconductivity) \cite{bochenek_2012} to metal-to-insulator transitions in PrRu$_4$P$_{12}$ and SmRu$_4$P$_{12}$ \cite{sekine_1997,sekine_1998}.
Rattling modes of the central $R$ atom have been observed in x-ray absorption fine structure measurements in the P- and Sb-skutterudites indicating their suitability for thermoelectric applications \cite{cao_2004}.

\begin{figure}[t]
\begin{center}
		{\includegraphics[width=.99\columnwidth]{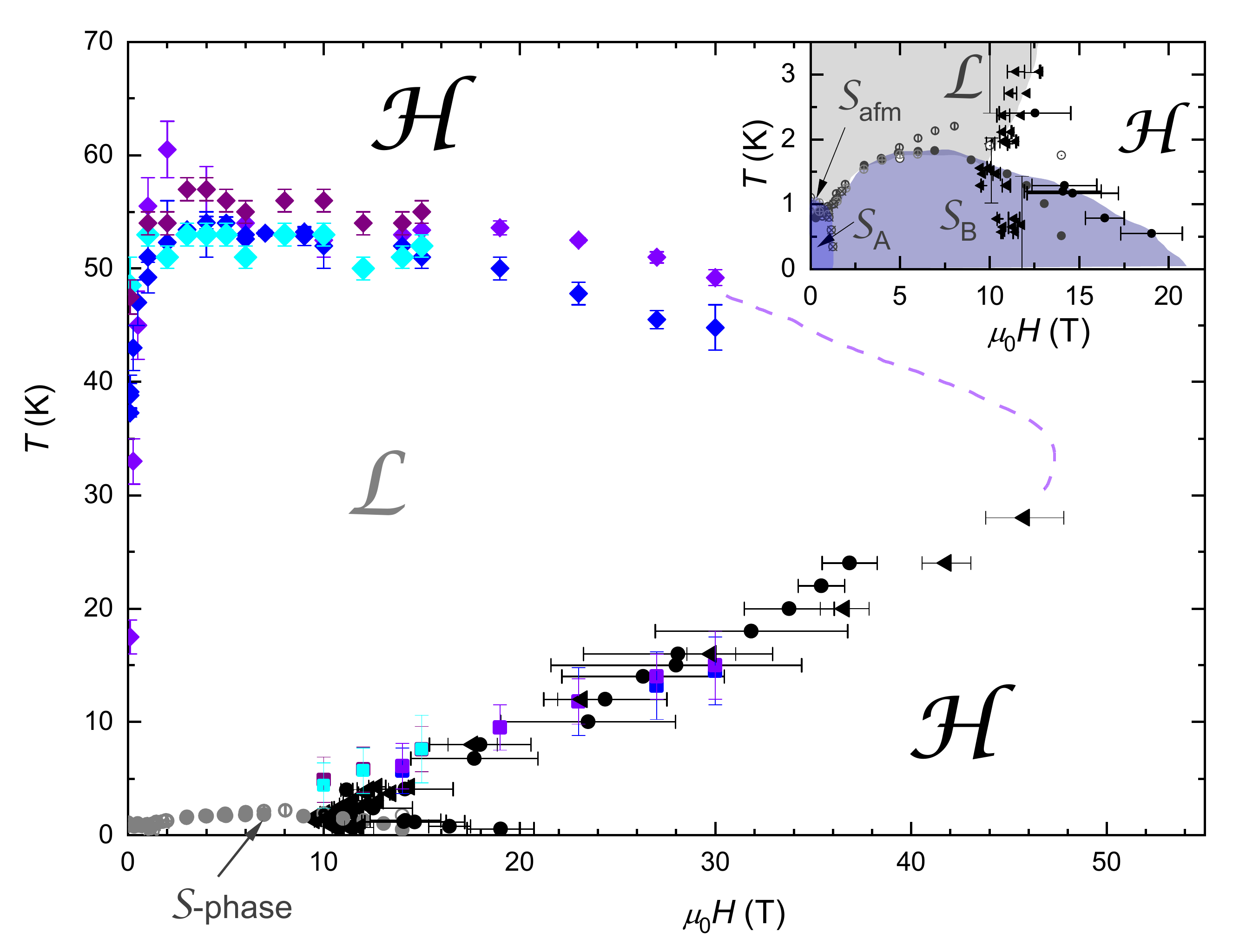}}
		\end{center}
   \caption[Ambient pressure phase diagram of \ceossb.]{
	Ambient pressure phase diagram of \ceossb as established in \cite{gotze_unusual_2020}. Different colours symbolize results from different crystals. Diamond-shaped and square symbols are from transport measurements, circles are from magnetostriction, and triangles from megahertz conductivity. See \cite{gotze_unusual_2020} for details. Grey symbols of the $\cal{S}$-phase boundary are from Refs.~\cite{tayama_2015,namiki_2003,sugawara_2005,rotundu_2006}.
	The inset focuses on the low-temperature $\cal{S}$-phase.
	}
	\label{fig:COS_phasedia}
\end{figure}

Most Ce-based skutterudites, however, show semimetallic, semiconducting or insulating behaviour \cite{meisner_1985,shirotani_1999,grandjean_1984}. \ceossb is a semimetal below about 50~K and exhibits a low-temperature ordered phase below 1~K \cite{bauer_2001,yogi_2005}.
In Refs.~\cite{ho_2016} and \cite{gotze_unusual_2020}, some of us have established the full ambient pressure $H$-$T$~phase diagram of \ceossb shown in Fig.~\ref{fig:COS_phasedia}.
A valence transition is found to occur between the metallic $\cal{H}$-phase and the semimetallic $\cal{L}$-phase, and between the $\cal{H}$-phase and the low-temperature $\cal{S}$-phase.
The boundary between these phases has an unusual, wedge-like shape and a gradient that alternates between negative and positive.
The $\cal{S}$-phase has in the past been identified as a spin-density wave phase, but evidence reported in Refs.~\cite{tayama_2015} and \cite{iwasa_magnetic_2008} by dilatometry, magnetization and neutron scattering suggests that there may in fact be three separate phases (see inset of Fig.~\ref{fig:COS_phasedia}): a magnetically ordered phase below 1~K and 1~T in the $H$-$T$ phase diagram that we call $\cal{S}_\mathrm{afm}$, a sub-phase of $\cal{S}_\mathrm{afm}$ here named $\cal{S}_\mathrm{A}$ which exists below $T$=900~mK, and a high-field ordered phase $\cal{S}_\mathrm{B}$ which is possibly an antiferroquadrupolar or charge-density wave phase \cite{tayama_2015}.
Quantum criticality close to the suppression of the $\cal{S}_\mathrm{B}$-phase towards zero temperature at approximately 20~T is associated with additional energy scales that cause the deviation of the phase boundary from the expected elliptical behaviour and also leads to a field-dependent increase of the effective quasiparticle masses as observed in quantum-oscillation experiments \cite{gotze_unusual_2020}. The Fermi surface (FS) of the $\cal{H}$-phase consists of a single, central spherical FS sheet theoretically predicted in Refs.~\cite{harima_2003} and \cite{yan_2012}, and experimentally confirmed by Ref.~\cite{ho_2016}.

Hydrostatic pressure can be used as a tuning parameter to drive phase transitions and band-structure properties of correlated electron systems. It can allow the investigation of quantum criticality \cite{helm_2020, singleton, Thomas_2020}, manipulate the band structure \cite{goddard_2002, shishido_2005} and lead to the emergence of new (superconducting) phases \cite{okada_2008,drozdov_2015} that would not be accessible at ambient pressure.
In contrast to alloying, the application of pressure not only preserves sample symmetry \cite{Clune_2020,Bangura_2007,Ghannadzadeh_2013} and introduces little disorder, but it is also continuous and reversible.
In this paper we present results on the pressure dependence of the phase boundaries of the high-temperature $\cal{H}$-to-$\cal{L}$ transition, of the $\cal{S}$-phase boundary and on the influence of pressure on the high-field Fermi surface and electronic band structure, and the corresponding effective masses.

\section{Methods}
\label{sec:methods}

Single crystals of \ceossb were grown as described in Refs.~\cite{ho_2016} and \cite{bauer_2001}. Several crystals from the same growth batch were used in the experiments presented in this paper.
Electrical transport measurements under pressure were conducted using piston-cylinder cells with Daphne 7373 or 7575 oil as the pressure medium. The pressure was determined \textit{in~situ} at low temperatures using either fluorescence of a ruby crystal or the superconducting critical temperature of tin. Measurements at the University of Warwick, UK took place in a $^4$He variable temperature insert situated in a 15~T superconducting magnet, while measurements at the NHMFL Tallahassee, USA were conducted in a 18~T superconducting magnet equipped with a $^3$He cryostat. Current was applied along the [100] axis.

Megahertz conductivity measurements (or radio-frequency techniques) allow for a contactless measurements of physical properties. The sample is coupled to the inductance or capacitance of the tank circuit of a $LCR$ circuit, and changes in, for example, magnetic or conductive properties are reflected in a change of resonance frequency \cite{ghannadzadeh_2011}.
For the megahertz conductivity measurements shown in this paper, the crystals were cleaved and small plates of $100~\mu\mathrm{m}$ by $100~\mu\mathrm{m}$ cross-section and $30-40~\mu\mathrm{m}$ thickness were loaded into a diamond-anvil cell which also contained a copper-wire coil with three or four turns that was part of the tunnel-diode oscillator (TDO) circuit.
The frequency change $\Delta f$ of a TDO circuit is proportional to the change in skin depth of the sample; in turn, the skin depth is proportional to the square root of the sample resistivity \cite{ghannadzadeh_2011,sakakibara_contactless_1989,coniglio_improvements_2010}. Maintaining this relationship places constraints on the crystal's lateral dimensions, which must be rather larger than (say $\approx$ 2~times) the skin depth to prevent the MHz oscillatory fields completely penetrating the sample. Once this size limit is exceeded, the change in frequency $\Delta f \propto -\Delta \rho$, where $\Delta \rho$ is the change in resistivity, as long as $\Delta f \ll f$, where $f$ is the TDO resonant frequency. Besides the signal coming from the sample, there is a background contribution to $\Delta f$ from the magnetoresistance of the copper coil which has to be subtracted.
The quantum oscillations observed in the field-dependent TDO signal of CeOs4Sb12 are Shubnikov-de Haas (SdH) oscillations of the resistivity due to the proportionality between $\Delta f$ and $-\Delta \rho$.

In the diamond-anvil cells, Nujol or a 1:4~methanol-ethanol mixture was used as a pressure medium.
The pressure was determined \textit{in~situ} at $T=2~\mathrm{K}$ from the fluorescence of a ruby crystal.
Experiments were conducted at ambient pressure and at different pressures up to 6.4~GPa.
These measurements took place at the NHMFL Tallahassee, USA in the 35~T and the 41.5~T resistive magnets.
Two pressure cells were installed on a low-temperature probe for each session and loaded into a $^3$He cryostat. One of the cells was mounted on a rotator platform that allows for angle-dependent measurements.

\section{Results}

\subsection{High-temperature resistivity up to 1.4~GPa}
\label{sec:highT}

\begin{figure}[t]
\begin{center}
		\subfigure{\includegraphics[width=.99\columnwidth]{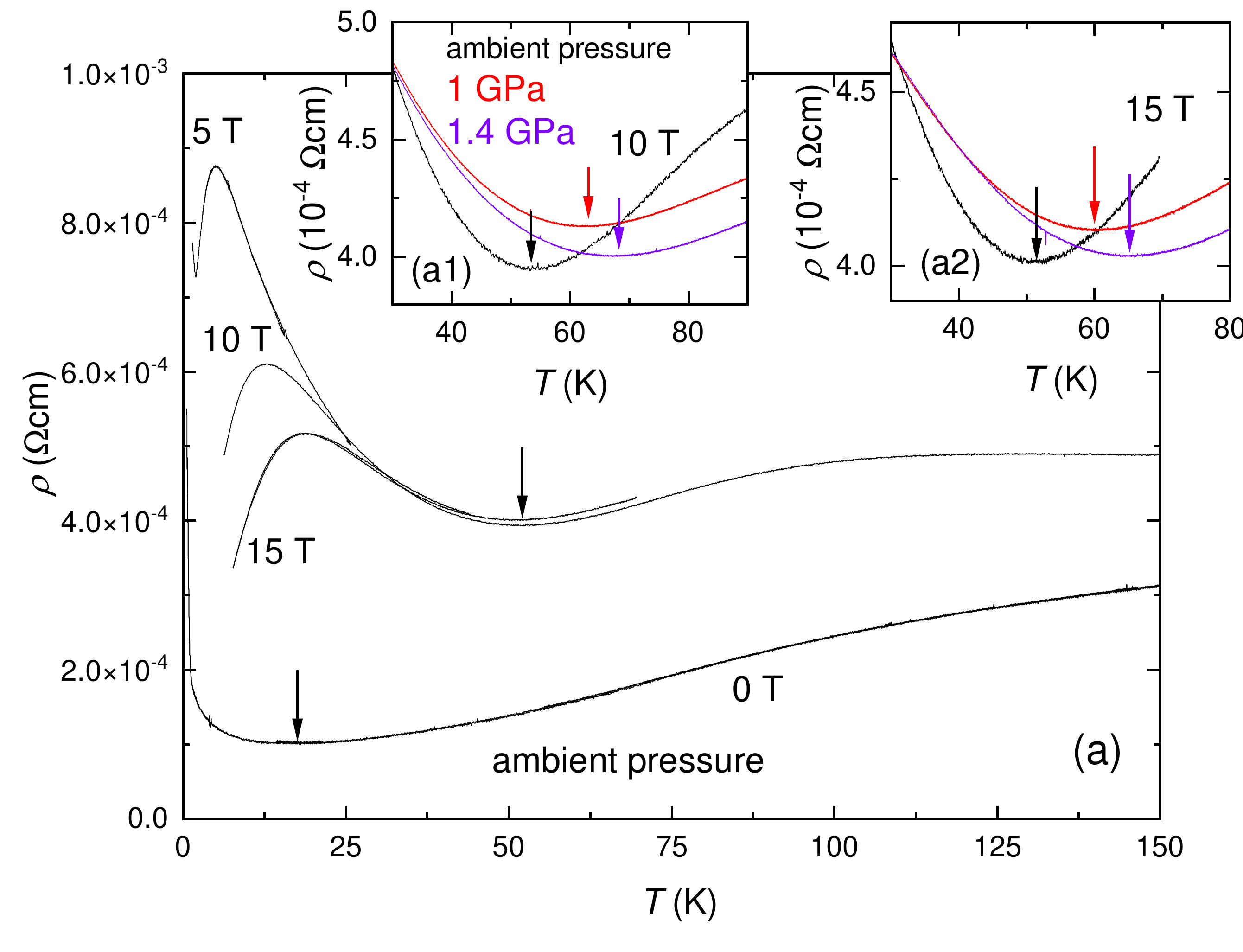}}\\
		\subfigure{\includegraphics[width=.99\columnwidth]{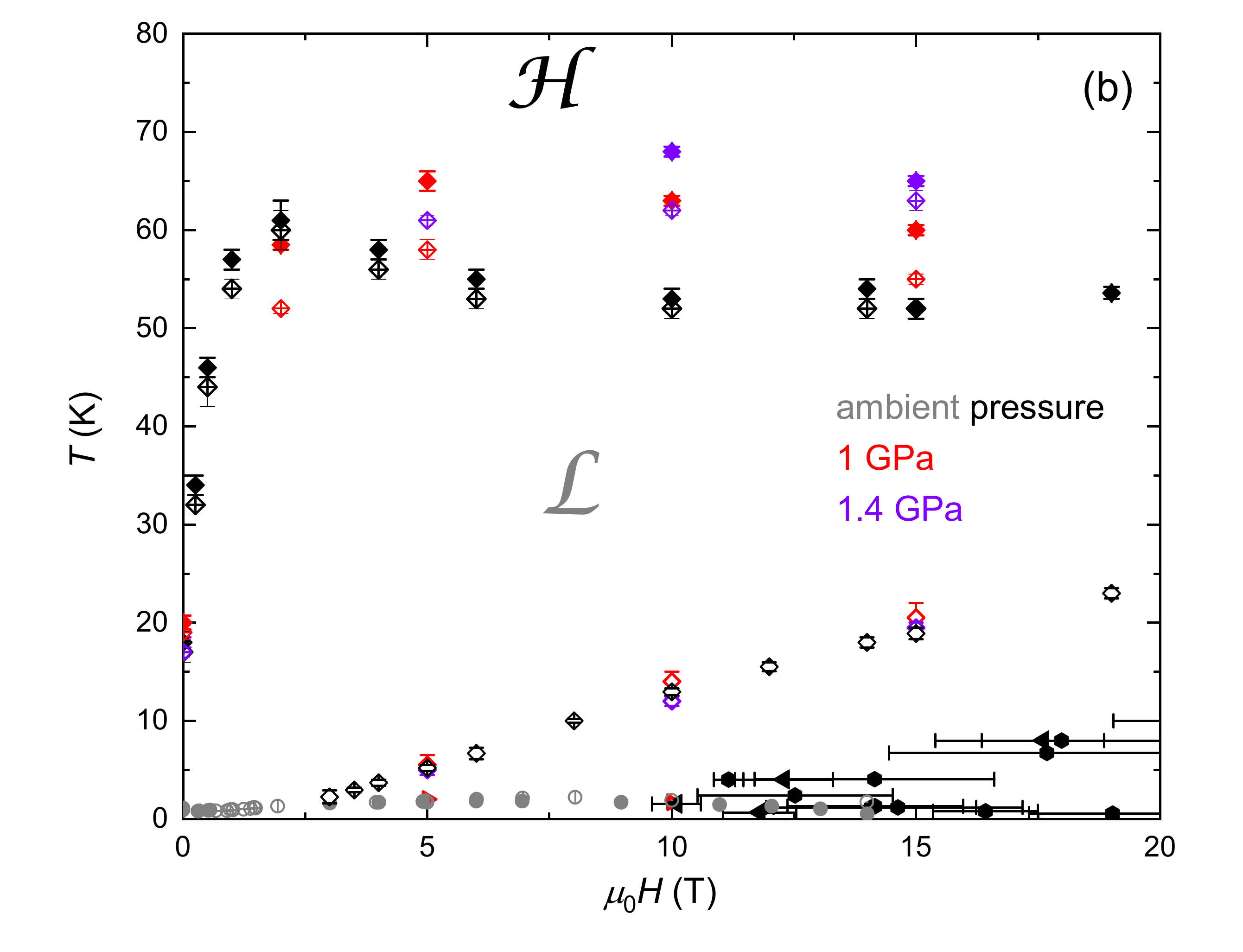}}
		\end{center}
   \caption[$\rho(T)$ at ambient pressure, 1 and 1.4~GPa and at different applied fields. Corresponding phase diagram of the high-temperature valence transition.]{
	(a) $\rho(T)$ at ambient pressure for zero field, 5, 10 and 15~T. The $\cal{H}$-to-$\cal{L}$ valence transition occuring at the local minimum is marked by arrows.
	Inset (a1) focuses on the $\rho(T)$ minimum at 10 T for ambient pressure (black), 1~GPa (red) and 1.4~GPa (purple).
	Inset (a2) shows $\rho(T)$ at 15~T; same pressures and colour coding as in (a1).
	(b) Phase diagram based on data in (a) and from Ref.~\cite{gotze_unusual_2020} illustrating the shift of the $\cal{H}$-to-$\cal{L}$ transition to higher temperatures with increasing pressure for fields of 5~T and above. Full (crossed) symbols were obtained upon cooling (warming), respectively; empty diamonds denote the temperature of the $\rho(T)$ maximum. Other ambient pressure points as in Fig.~\ref{fig:COS_phasedia}.
	}
	\label{fig:COS_highT}
\end{figure}

The main panel of Fig.~\ref{fig:COS_highT}\,(a) shows the temperature dependence of the resistivity $\rho(T)$ of \ceossb at ambient pressure for zero field, 5, 10 and 15~T.
The resistivity initially decreases for decreasing temperatures and goes through a minimum at the temperature $T_\mathrm{min}$ (marked by arrows) for all applied fields.
The zero-field $\rho(T)$ then increases exponentially as the temperature is lowered, whereas for applied fields $\geq 3~\mathrm{0}T$ the resistivity increase is less steep and $\rho(T)$ goes through a local maximum upon further cooling.
The minimum marks the high-temperature valence transition between the $\cal{H}$- and the $\cal{L}$-phase while the local maximum (empty diamonds in Fig.~\ref{fig:COS_highT}\,(b)) was established to be a precursor of the field-induced $\cal{L}$-to-$\cal{H}$ transition \cite{gotze_unusual_2020}: 
As the material crosses over from insulating to metallic behaviour at lower temperatures for fields above 3~T, a maximum in $\rho(T)$ is inevitable.
However, only for fields greater than $\approx 10~\mathrm{T}$ is the $\rho(T)\propto T^2$ dependence that is expected for metals observed; below this field the $\cal{S}$-phase intervenes.
Therefore, the maximum in $\rho(T)$ at $T_\mathrm{max}$ should be seen as a precursor to the restoration of full metallic behaviour in the low-temperature high-field $\cal{H}$-phase.

Application of pressure moves the $\cal{H}$-to-$\cal{L}$ transition and therefore the $\rho(T)$ minimum to higher temperatures as shown in insets (a1) and (a2) of Fig.~\ref{fig:COS_highT} for 10 and 15~T.
We have also observed a shift of $T_\mathrm{min}$ at 0 and 2~T but it does not continuously move to higher temperatures as it does for fields $\geq 5~\mathrm{T}$.
The phase diagram in Fig.~\ref{fig:COS_highT}\,(b) illustrates these results and also the hysteresis between data recorded upon warming (crossed symbols) and cooling (full symbols).
Hysteretic behaviour at a Ce valence transition has been observed before \cite{thompson_1983,gignoux_1985,drymiotis_2005} and has been discussed in detail in \cite{ho_2016} and \cite{gotze_unusual_2020}. Note that the position of the maximum (empty diamonds) is not hysteretic and hardly affected by pressure. 
This feature is, as mentioned above, a precursor to the field-induced $\cal{L}$-to-$\cal{H}$ transition and will be discussed later in context with the other high-field phenomena in \ceossb.

\subsection{Suppression of the $\cal{S}$-phase}
\label{sec:SDW}

\begin{figure}[t]
\begin{center}
		\subfigure{\includegraphics[width=.99\columnwidth]{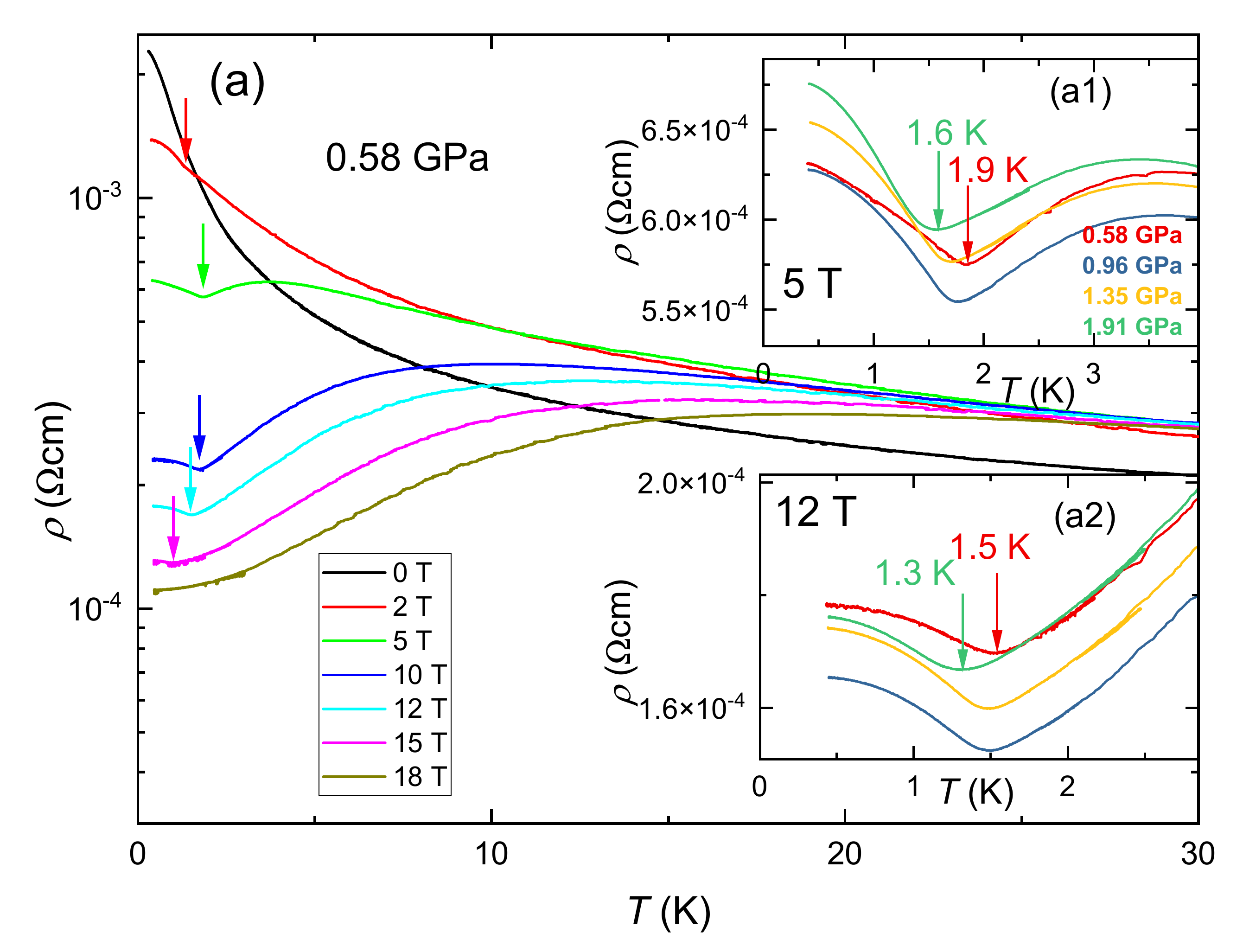}}\\
		\subfigure{\includegraphics[width=.99\columnwidth]{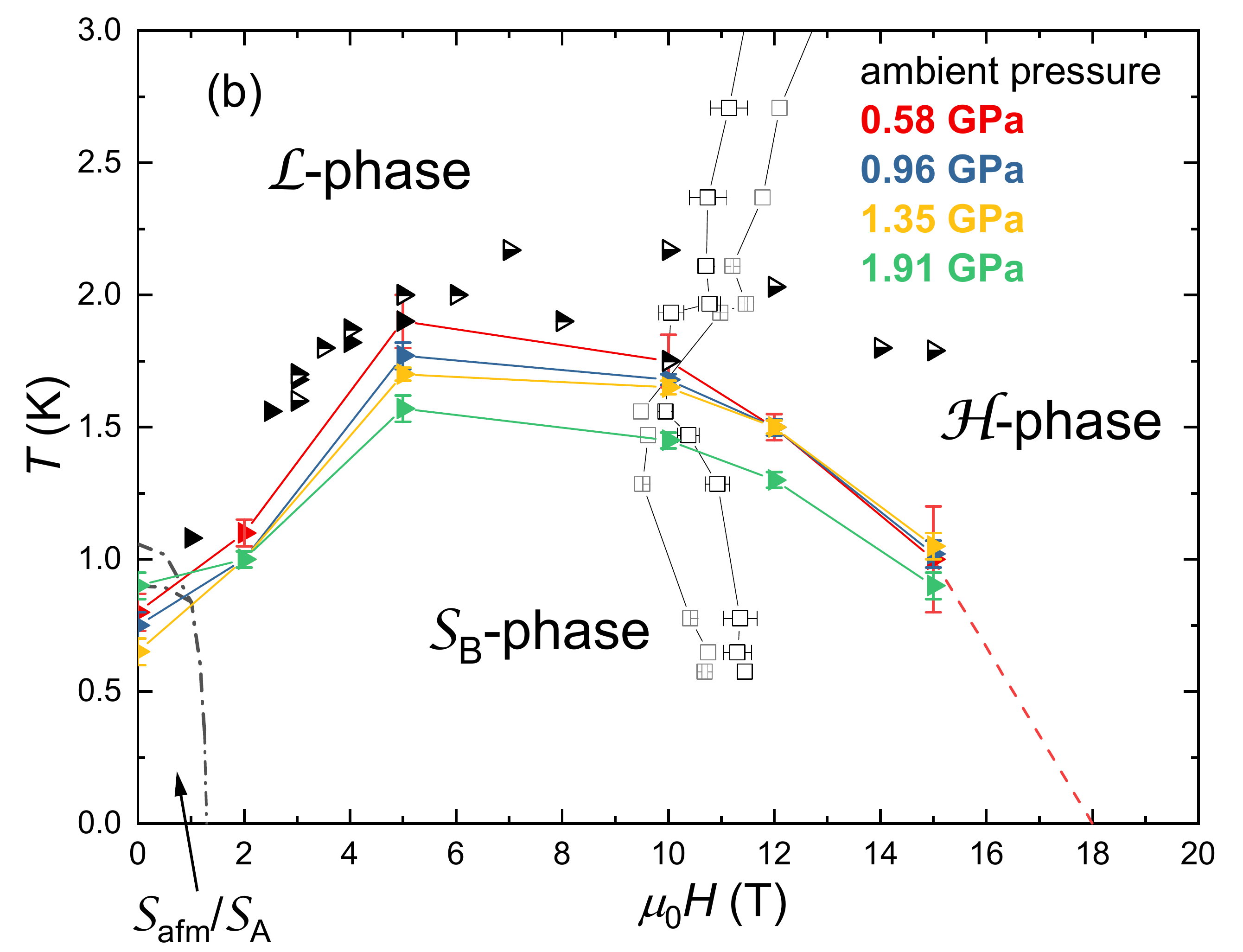}}
		\end{center}
   \caption[$\rho(T)$ at different pressures between 0.58 and 1.91~GPa and corresponding phase diagram of the $\cal{S}$-phase.]{
	(a) Cool down curves $\rho(T)$ at 0.58~GPa for different fixed magnetic fields. The transition from the $\cal{L}$-phase to the $\cal{S}$-phase is marked by arrows.
	Inset~(a1): Focus on $\rho(T)$ around the $\cal{S}$-phase transition at 5~T for different pressures.
	Inset~(a2): Same as (a1) but for an applied field of 12~T.
	(b) Phase diagram of the $\cal{S}$-phase for different pressures. All full symbols are based on measurements on the same crystal, half-full and empty symbols are results from different crystals. Triangles: Position of the kink in $\rho(T)$ from (a) and Ref.~\cite{gotze_unusual_2020}; squares: field-induced transition in megahertz conductivity from \cite{ho_2016}, grey: rising field, black: falling field. The dashed line is a guide to the eye considering that the kink in $\rho(T)$ has not been observed at 18~T for all pressures. The dash-dot line at lower fields that separates the sub-phases of the $\cal{S}$-phase is based on Ref.~\cite{tayama_2015}.
	}
	\label{fig:COS_SDW}
\end{figure}

The transition into the $\cal{S}$-phase is indicated by a kink in the $\rho(T)$ curve at a temperature $T_{\cal{S}}$ marking either a strong change (0--3~T) or a reversal of slope ($\mu_0 H > 3~\mathrm{T}$) \cite{sugawara_high_2004}. The main panel of Fig.~\ref{fig:COS_SDW} shows the development of the position of this kink (marked by arrows) for different applied fields at 0.58~GPa. As for ambient pressure conditions \cite{namiki_2003,sugawara_2005}, $T_{\cal{S}}$ initially increases with increasing fields before being suppressed for fields above 10~T. The transition was observed for 15~T but not for 18~T. The application of higher pressures moves $T_{\cal{S}}$ to lower temperatures for all fields above 2~T. Insets (a1) and (a2) of Fig.~\ref{fig:COS_SDW} illustrate this behaviour for 5 and 12~T showing a suppression of $T_{\cal{S}}$ by 300 and 200~mK, respectively, at 1.91~GPa. For the zero-field and 2~T measurements, no or only a small suppression of the $\cal{S}$-phase transition temperature is observed. As was shown in \cite{sato_transport_2003} and \cite{tayama_2015} the $\cal{S}$-phase consists of at least two subphases with the low-field antiferromagnetic phase being suppressed above 1~T. This could explain the different behaviour of $T_{\cal{S}}$ close to this phase boundary as compared to higher fields.
The phase diagram in Fig.~\ref{fig:COS_SDW}\,(b) summarizes the development of the $\cal{S}$-phase up to 1.91~GPa. We find the strongest suppression of $T_{\cal{S}}$ to occur between 5 and 12~T. From the minor effect the pressure has on the transition at 15~T, we do not expect the critical field at which $T_{\cal{S}}$ is suppressed to zero to move significantly in this pressure range.

\subsection{Quantum oscillations and effective masses}
\label{sec:QO}

\begin{figure}[t]
\begin{center}
		\subfigure{\includegraphics[width=.99\columnwidth]{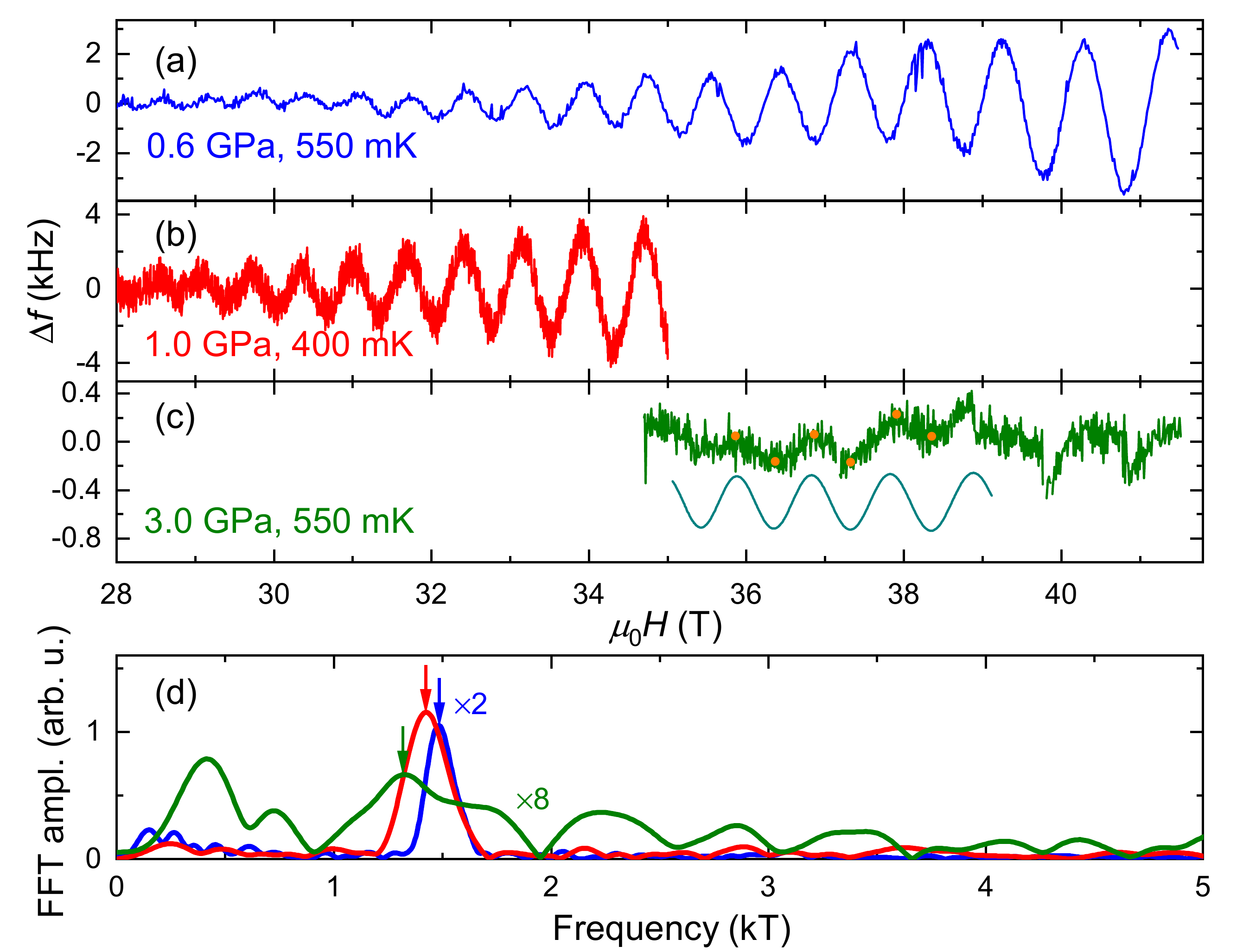}}\\
		\subfigure{\includegraphics[width=.99\columnwidth]{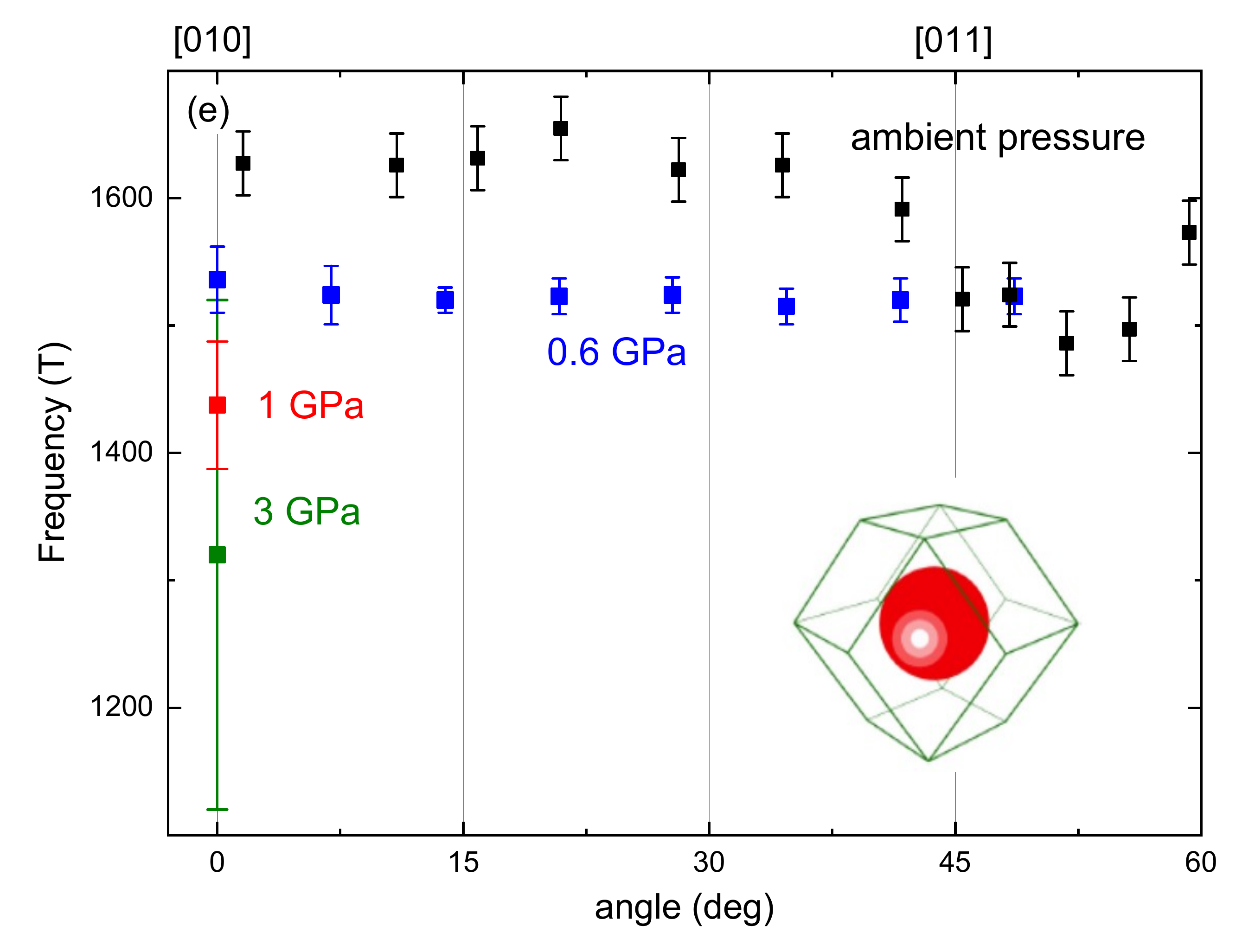}} 
		\end{center}
   \caption[Examples for quantum oscillations at 0.6, 1.0 and 3.0~GPa and angular dependence of dHvA frequencies.]{
	(a)-(c) Examples of the oscillating part of the TDO signal at 0.6, 1.0 and 3.0~GPa, respectively; a 3rd or 4th order polynomial has been subtracted from the measured data. In (c) we have highlighted minima and maxima of the oscillating experimental data by orange dots. The smooth curve is a simulation of quantum oscillations with frequency 1394~T.
	(d) Frequency spectrum obtained by fast Fourier transformation (FFT) of the oscillating data. Colour coding as above.
	(e) Main panel: Angular dependence of the quantum oscillation frequency at ambient pressure and at 0.6~GPa and frequencies at $\mu_0 H\parallel[010]$ for 1.0 and 3.0~GPa. Calculated FS of the $\cal{H}$-phase (from Ref.~\cite{ho_2016}).
	}
	\label{fig:COS_angledep_osc}
\end{figure}

Quantum oscillations under pressure have been observed in TDO measurements for fields above 25~T for 0.6 and 1.0~GPa and above 34.7~T at 3.0~GPa.
Figs.~\ref{fig:COS_angledep_osc}\,(a)--(c) show the oscillating part of the TDO signal at the lowest measured temperature after the subtraction of a polynomial background at 0.6, 1.0 and 3.0~GPa, respectively.
As the pressure increases, the signal quality is reduced, probably due to the hydrostaticity limit of the Nujol, and weak SdH oscillations at 3.0~GPa were observed only above 34.7~T at 550~mK. 
We have highlighted the maxima and minima of the 3.0~GPa oscillating signal in Fig.~\ref{fig:COS_angledep_osc}\,(c) with orange dots and we have calculated the oscillation frequency based on the position of the marked minima and maxima, obtaining 1.4(2)~kT.
Simulated quantum oscillations of this frequency are included in Fig.~\ref{fig:COS_angledep_osc}\,(c) to facilitate the identification of the oscillations.
The frequency spectra at 0.6, 1.0 and 3.0 GPa obtained by fast Fourier transform (FFT) are shown in Fig.~\ref{fig:COS_angledep_osc}\,(d). The frequency of the oscillations decreases with increasing pressure from 1.49--1.63~kT (depending on the orientation of the sample in the field) at ambient pressure \cite{ho_2016} to 1.3(2)~kT at 3.0~GPa.
The frequency at 3.0~GPa obtained by FFT (1.3(2)~kT) is the same within error bars as the one calculated from minima and maxima of the signal (1.4(2) T). However, both are lower than the frequency at 1.0~GPa. In Fig.~\ref{fig:COS_angledep_osc}\,(e) we show the frequencies obtained by FFT.
A comparison between the angular dependence of the SdH oscillations at ambient pressure (inset of Fig.~\ref{fig:COS_angledep_osc}\,(e)) and 0.6~GPa (main panel of Fig.~\ref{fig:COS_angledep_osc}\,(e)) shows that the angular variation of the frequency diminishes under pressure. That means that the $\cal{H}$-phase FS of \ceossb becomes slightly smaller and more spherical when applying pressure.
This result is rather surprising -- typically a simple contraction of lattice parameters by pressure would lead to an enlargement of the FS; the observed decrease in FS size is, therefore, an indication of more profound changes in the electronic structure of \ceossb. Prior ambient pressure experiments have shown no variation of the size of the $\cal{H}$-phase FS between different samples other than the angular variation shown in Fig.~\ref{fig:COS_angledep_osc}\,(e).
(Note that a persistence to high fields of the semimetallic $\cal{L}$-phase FS would not account for our observations at high pressure. This FS has been calculated in \cite{yan_2012}, and consists of ellipsoidal pockets in the center (hole) and around the H-point (electron) of the Brillouin zone, with cross-sectional areas corresponding to quantum-oscillation frequencies of around 20~T. This compensated, low-carrier-density system is in qualitative agreement with Hall-effect data measured in the $\cal{L}$-phase \cite{Hall}.)

\begin{figure}[t]
\begin{center}
		\subfigure{\includegraphics[width=.9\columnwidth]{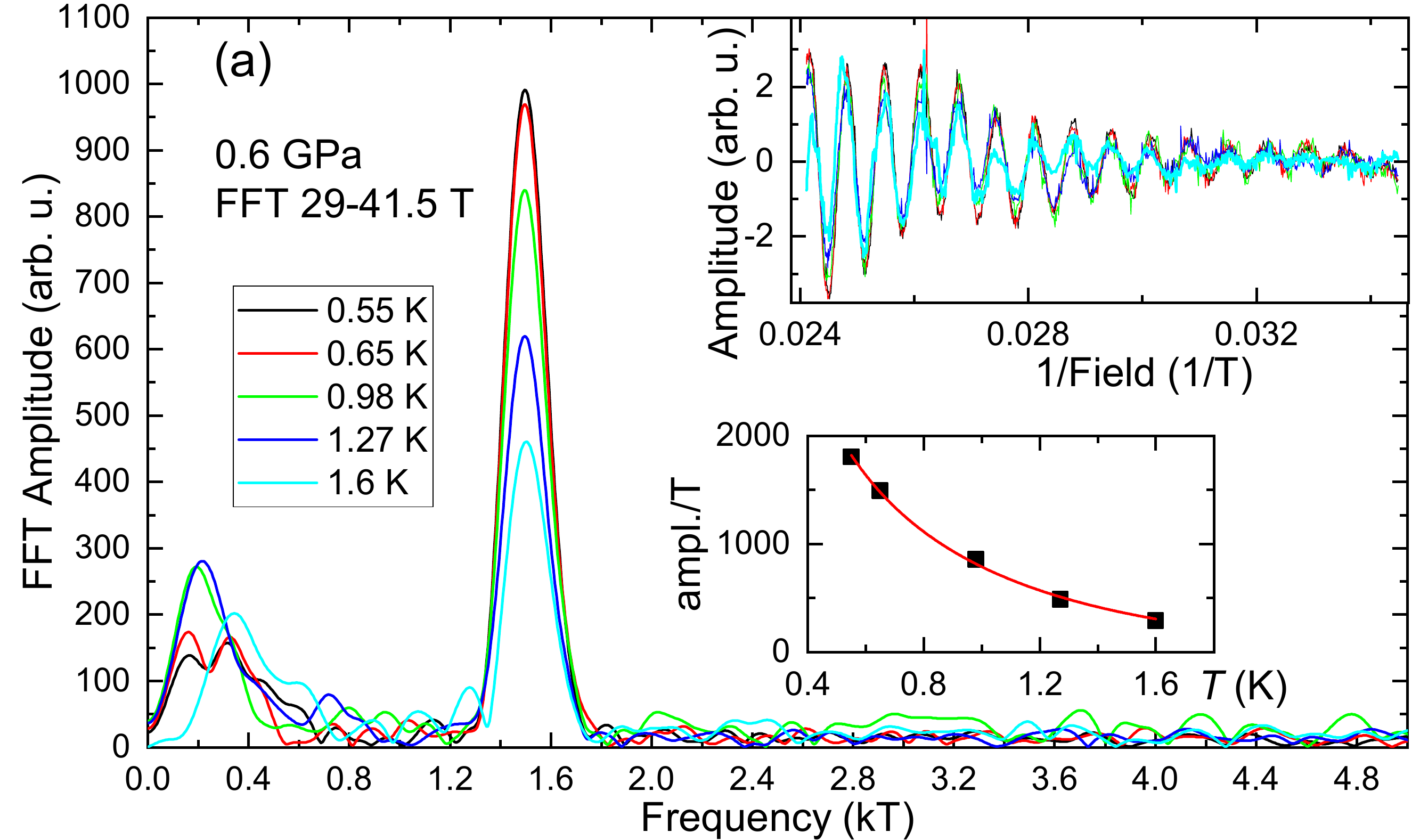}}\\
		\subfigure{\includegraphics[width=.9\columnwidth]{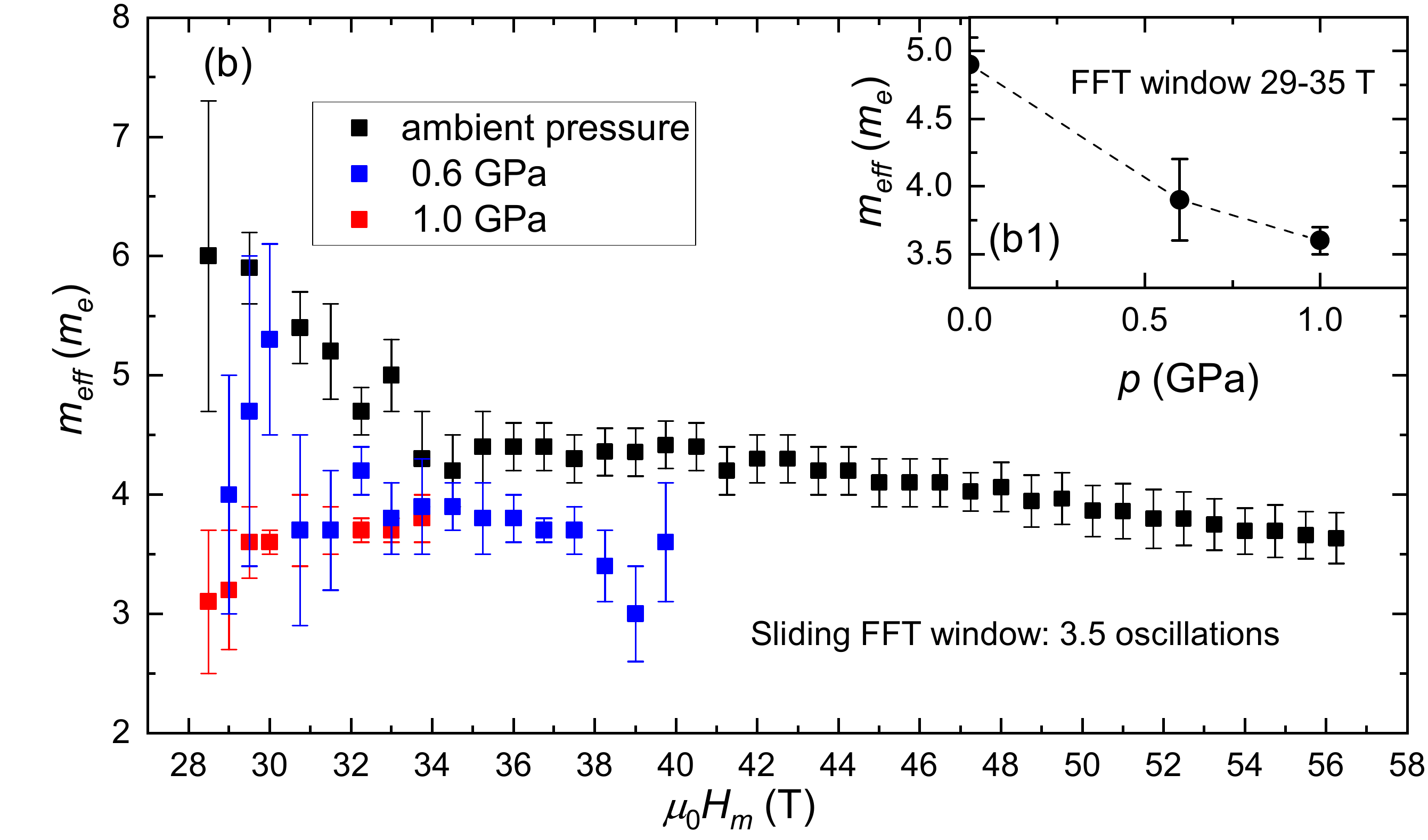}}\\
		\end{center}
   \caption[Temperature dependence of quantum oscillation frequency at 0.6~GPa and field-dependence of effective masses at different pressures.]{
	(a) Main panel: Frequency spectrum at 0.6~GPa at several temperatures between 0.55 and 1.6~K. Upper inset: Oscillating part of the signal plotted as $1/\mu_0 H$ for the same temperatures as in the main panel. Bottom inset: FFT amplitudes divided by temperature as a function of temperature and fit line determined by use of the Lifshitz-Kosevich formula yielding $m_\mathit{eff}= 3.5(2) m_e$ with $m_\mathrm{e}$ being the rest electron mass.
	(b) Field dependence of the effective masses for ambient pressure (pulsed-field, black), 0.6 (blue) and 1.0~GPa (red symbols).
	Inset (b1): Effective mass for the FFT range 29 to 35~T as a function of pressure. The dashed line is a guide to the eye.
	}
	\label{fig:COS_meff}
\end{figure}

The effective masses of quasiparticles can be determined from the suppression of the quantum-oscillation amplitude with increasing temperature by fitting to the well-known Lifshitz-Kosevich (LK) formula \cite{shoenberg}. The main panel of Fig.~\ref{fig:COS_meff}\,(a) shows the frequency spectrum of the quantum oscillations of \ceossb at 0.6~GPa for temperatures between 0.55 and 1.6~K. The upper inset shows the oscillating part of the signal at the same temperatures as the main panel while the lower inset depicts the $T-$dependence of the Fourier amplitude divided by $T$ along with a fit to the LK function. This fit, employing a wide field window for the Fourier transforms [here $29-41.5~$T ] that covers the region over which oscillations are observed, yields an effective mass value of  
$m_{\rm eff}/m_{\rm e} = 3.5 \pm 0.2$.
At ambient pressure, pulsed-field measurements in Ref.~\cite{gotze_unusual_2020} showed that the effective mass is in fact field-dependent, increasing with decreasing $H$. In order to test whether this field dependence persists at applied pressures, we follow a similar procedure to analyse the temperature dependence of the quantum oscillations for the 0.6 and 1.0~GPa data here. We divide the field range over which oscillations are observed into smaller windows covering $1/467~{\rm T}^{-1}$ in inverse field (containing roughly 3.5 oscillations) and determine the effective mass for each of those windows using fits to the LK function. The results are shown versus $\mu_0H_{\rm m}$, 1/(the mean inverse field) of each window, in Fig.~\ref{fig:COS_meff}\,(b), which also includes ambient-pressure data from~\cite{gotze_unusual_2020}. At ambient pressure (black symbols) $m_{\rm eff}$ increases slightly for decreasing field between 56 and 35~T but more strongly below 35~T, when the system approaches the quantum critical point (QCP) associated with the suppression of the $\cal{S}$-phase.
This increase was associated with strong quantum fluctuations close to the QCP.
At 0.6 (blue symbols) and 1.0~GPa (red symbols) we do not observe such a strong field dependence of the effective mass.
For 0.6~GPa the effective mass is approximately constant at 3.8~$m_\mathrm{e}$ between 37.5 and 30.75~T with some lower masses between 3.0 and 3.7~$m_\mathrm{e}$ at higher fields. At lower fields, we find higher masses between 4.0 and 5.3~$m_\mathrm{e}$ but as the signal decreases in strength at lower fields, these results are accompanied by larger error bars.
At 1.0~GPa we observe an almost constant effective mass of 3.7--3.6~$m_\mathrm{e}$ between 33.75 and 29.5~T. $m_\mathrm{eff}$ decreases for lower fields but again we have to consider the larger error bars for these values.

The absence of a strong enhancement of $m_{\rm eff}$ towards lower fields is confirmed by regarding the effective mass that was determined for a field window 29--35~T for each pressure. Inset~(b1) of Fig.~\ref{fig:COS_meff}\,(b) shows the development of $m_\mathrm{eff}$ as a function of pressure, clearly illustrating a decrease in effective mass with increasing pressure.


\section{Discussion}

\subsection*{High-temperature valence transition and sample dependence}

We observe an increase in the position of the $\rho(T)$ minimum at $T_\mathrm{min}$ that marks the high-temperature $\cal{H}$-to-$\cal{L}$ valence transition with increasing pressure for fields $\geq 5~\mathrm{T}$.
A similar increase with pressure in zero field has been observed previously by Ref.~\cite{hedo_pressure_2003}. However, in contrast to the study in Ref.~\cite{hedo_pressure_2003}, we only observe a systematic increase of $T_\mathrm{min}$ with pressure for fields above 5~T; for lower fields we do not observe a clear trend and, in some cases, $T_\mathrm{min}$ decreases with pressure.
This behaviour might be related to a strong sample dependence of $T_\mathrm{min}$. 
As can be seen in the ambient pressure phase diagram in Fig.~\ref{fig:COS_phasedia} and as was discussed in detail in Ref.~\cite{gotze_unusual_2020}, the critical temperature of the $\cal{H}$-to-$\cal{L}$ transition in \ceossb is sample dependent, in particular at zero field and low applied fields, even within the same growth batch. This explains the different $T_\mathrm{min}$ in zero field in Ref.~\cite{hedo_pressure_2003} and in our study.
In Ref.~\cite{rotundu_exotic_2007} Ce vacancies were identified as a source for sample dependence for the transition to the $\cal{S}$-phase~\cite{iwasa_magnetic_2008,namiki_2003,ito_microscopic_2010}. However, in Ref.~\cite{gotze_unusual_2020} we argue that inclusions of very small amounts of elemental atoms, such as Sb, can cause local uniaxial stress that leads to the presence of tiny domains which are topologically protected~\cite{gotze_unusual_2020,yan_2012}; these will strongly influence the transport properties of \ceossb at zero and low fields \cite{gotze_unusual_2020} and cause the above mentioned variation of $T_\mathrm{min}$.
There is experimental evidence that topologically protected states are sensitive to relatively small magnetic fields~\cite{he_2011,liu_2012,yang_2013,zheng_2016,rebar_2019}, which may explain the large positive gradient of the $\cal{H}$-to-$\cal{L}$ boundary in the ambient pressure $H$-$T$ phase diagram below 3~T and why the sample dependence of $T_\mathrm{min}$ is clearly reduced for applied fields above 3~T. Fig.~\ref{fig:COS_phasedia} shows that for fields higher than 3~T the variation of $T_\mathrm{min}$ between different samples is reduced to a few kelvin. 

We will now put our results in context with the above mentioned zero-field result of Ref.~\cite{hedo_pressure_2003} and the pressure dependence of the valence transition of elemental cerium. We observe the biggest shift of the $\cal{H}$-to-$\cal{L}$ transition in \ceossb at 10~T with a rate of 10.7~K/GPa. Hedo \textit{et~al}. report a linear increase of $T_\mathrm{min}$ in zero field of 4.25~K/GPa up to 8~GPa (where the $\rho(T)$ minimum is still observed) which is comparable to the shift of 5~K/GPa that we observe at 5~T. In elemental cerium, the $\gamma-\alpha$ valence transition is shifted from 120~K at ambient pressure to a critical end point at 550~K and 1.9~GPa, where the characteristic volume collapse vanishes, corresponding to 226~K/GPa \cite{gschneider}.
This difference in shift rate between \ceossb and elemental cerium is a consequence of the small ratio of cerium to other elements in \ceossb (1:16). The involvement of cerium in the bonding in \ceossb is, therefore, greatly reduced, which is consequently reflected in the rather smooth structural transition observed in magnetostriction \cite{gotze_unusual_2020}. As pressure $p$ and volume $V$ are conjugate variables in thermodynamics, any dramatic shift in volume accompanying a phase transition (such as in elemental Ce) suggests an extreme sensitivity to pressure. Conversely, the reduced change in volume that accompanies the valence transition in \ceossb suggests a much smaller shift with pressure.
Another aspect to consider is that the phase boundary of the valence transition in \ceossb has been altered by the sensitivity of the ground states to quantum fluctuations and to the proximity to a topological semimetallic phase \cite{gotze_unusual_2020}. The elliptical shape of the phase boundary in cerium, following $H^2\propto T^2$, is transformed to a more wedge-like boundary of the $\cal{L}$-phase and the strong correlations responsible for the deviation of the phase boundary could help the presence of two different valence configurations to survive to higher pressures.

\subsection*{Fermi surface}

\begin{figure}[t]
\begin{center}
		\includegraphics[width=.99\columnwidth]{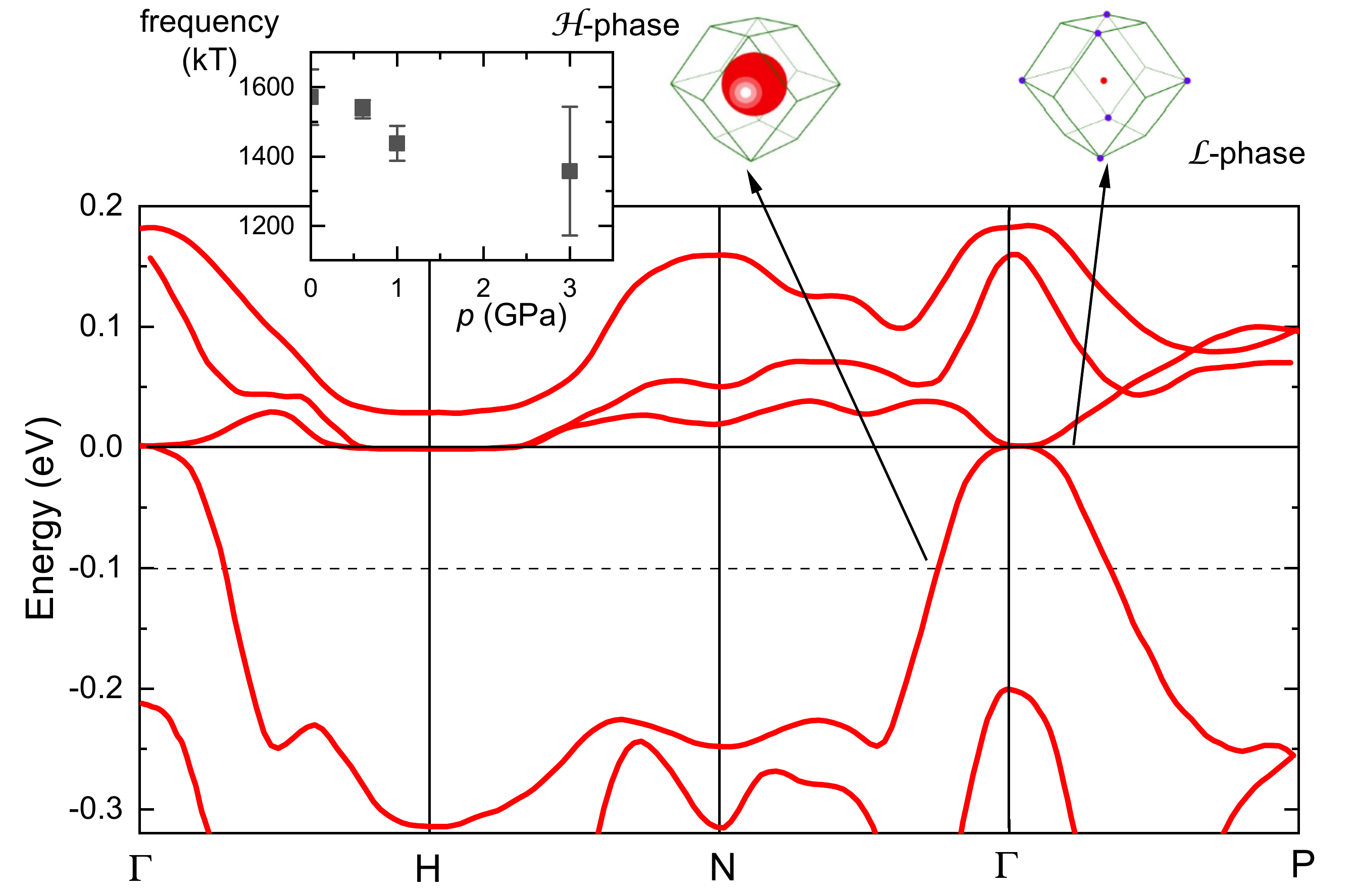}		
		\end{center}
   \caption[Band dispersion and quantum oscillation frequency as a function of pressure.]{Main panel: Band dispersion of \ceossb adapted from \cite{yan_2012}. $E=0$ corresponds to the Fermi level $E_\mathrm{F}$ in the $\cal{L}$-phase, whereas the dashed line at $E=-0.1~\mathrm{eV}$ roughly corresponds to $E_\mathrm{F}$ in the $\cal{H}$-phase. The arrows point to representations of the corresponding FSs (from Ref.~\cite{ho_2016}).
		Inset: quantum oscillation frequency for $\mu_0 H\parallel[010]$ as a function of pressure showing a 14\% reduction in the size of the $\cal{H}$-phase FS.
	}
	\label{fig:COS_band}
\end{figure}

The size of the $\cal{H}$-phase FS decreases with increasing pressure as shown in the inset of Fig.~\ref{fig:COS_band}. As we stated in section~\ref{sec:QO}, this is an unexpected result because the consequence of a simple contraction of lattice parameters with pressure would be a larger rather than a smaller FS. We surmise that the application of pressure leads to a change of the effective Ce valence in the $\cal{H}$-phase which shifts the Fermi energy.
The band dispersion of \ceossb in the main panel of Fig.~\ref{fig:COS_band} (adapted from~\cite{yan_2012}) shows the different Fermi-energy levels of the $\cal{L}$-phase ($E=0$) and the $\cal{H}$-phase (dashed line at $E=-0.1~\mathrm{eV}$) and the corresponding calculated FSs. A decreasing FS size in the $\cal{H}$-phase could be accounted for by a small reduction in the size of the shift of $E_\mathrm{F}$ towards -0.1~eV that occurs at the valence transition, possibly caused by changes in the localisation process of the Ce 4$f$-electrons at the $\cal{L}$-to-$\cal{H}$ valence transition.

\subsection*{Phase diagram and QCP}

\sloppypar The different phase boundaries of the \ceossb $H$-$T$-phase diagram were found to be shifted by pressure at very different rates. While the high-temperature Ce valence transition between the $\cal{H}$- and the $\cal{L}$-phase was moved towards higher temperatures for applied fields with rates of the order of 10~K/GPa, the low-temperature $\cal{S}$-phase was suppressed by about 0.23~K/GPa and still extends to at least 15~T at 1.9~GPa.
These two transitions are governed by very different energy scales: the $\cal{H}$-to-$\cal{L}$ valence transition is accompanied by a FS reconstruction and a change in $f$-electron character, whereas the $\cal{L}$-to-$\cal{S}$ phase transition is a transition into a weakly-ordered state \cite{tayama_2015}. Both the entropy change (0.05$R\ln2$ \cite{bauer_2001,namiki_2003}) and the change in magnetostriction \cite{tayama_2015} at this latter phase transition are small, suggesting that neither a FS reconstruction occurs, nor a significant change in lattice parameters and charge-carrier distribution. By analogy with the arguments in the previous section, the absence of a strong volume change upon the $\cal{L}$-to-$\cal{S}$ transition will make it less sensitive to pressure.

When evaluating the pressure dependence of the field-induced $\cal{L}$-to-$\cal{H}$ transition, we can look at the pressure dependence of the $\rho(T)$ maximum at $T_\mathrm{max}$ for fixed fields above 3~T, as this feature was identified as a precursor to the $\cal{L}$-to-$\cal{H}$ transition \cite{gotze_unusual_2020}. $T_\mathrm{max}$ does not experience a continuous suppression for all fields but for 5 and 10~T at 1.91~GPa it is lowered by 0.3 and 0.6~K, respectively, compared to the value at 0.58~GPa, which is very similar to the suppression of the $\cal{S}$-phase.
Since the position of $T_\mathrm{max}$ and the $\cal{L}$-to-$\cal{H}$ transition are strongly coupled, this behaviour implies that the field-induced $\cal{L}$-to-$\cal{H}$ transition is not significantly altered by about 2~GPa.
This conclusion is further supported by the fact that at ambient pressure the $\cal{L}$-to-$\cal{H}$ transition shares its low temperature boundary with the $\cal{S}$-phase, and both $T_\mathrm{max}$ and $T_{\cal{S}}$ are only marginally affected by the pressures used here.

The measurements presented in Ref.~\cite{gotze_unusual_2020} suggest that the high temperature $\cal{H}$-to-$\cal{L}$ transition and the field-induced $\cal{L}$-to-$\cal{H}$ transition are connected by the same phase boundary (see also Fig.~\ref{fig:COS_phasedia}). Yet the high-$T$ portion of this phase boundary is affected more strongly by pressure than the low-$T$/high-$H$ portion. We have already discussed that the field-induced $\cal{L}$-to-$\cal{H}$ transition shares its low-$T$ phase boundary with the $\cal{S}$-phase which will influence its reaction to pressure. Another important energy scale is the $-TS$ term of the free energy ($S$ being the entropy) that determines the field dependence of the valence transition in Ce and Yb compounds \cite{dzero_2000}. Higher temperatures emphasize the importance of this term, and so the pressure dependence will be significantly different.

This leaves us with the question of why the increase in effective mass towards lower field is dampened or absent at higher pressures, when the QCP associated with the field-induced suppression of the $\cal{S}$-phase is not moved significantly to lower fields.
We have observed that the strongest change in $T_{\cal{S}}$ with pressure occurs between 5 and 12~T, while for higher and lower fields the suppression of $T_{\cal{S}}$ is more subtle. This makes the $\cal{S}$-phase dome more flat, possibly changing the character of the $\cal{S}$-phase boundary with pressure which, in turn, promotes thermal over quantum fluctuations at the critical field of the phase suppression.
Another possibility is that the system in the $\cal{H}$-phase is pushed through a ``hidden'' QCP located somewhere in the $p$–$H$ phase diagram. At ambient pressure we pass close to this QCP, resulting in the observed enhancement of the effective mass, whereas at higher pressures of 0.6 and 1.0~GPa away from this QCP, the relative absence of critical fluctuations would explain the largely field independent effective masses. This type of quantum criticality has been observed in the heavy-fermion system CeCoIn$_5$ and is associated with a delocalization of 4$f$ electrons but instead of symmetry breaking, for example in the form of magnetic ordering, it is accompanied by a fractionalization of spin and charge \cite{cecoin5}.


\section{Summary}

We have conducted electrical transport and megahertz conductivity TDO measurements under pressures of up to 3~GPa and in magnetic fields up to 41~T on different \ceossb single crystals. Our results show a pressure-induced shift of the high-temperature Ce valence transition towards higher temperatures, similar to the pressure effect on the cerium $\gamma-\alpha$ transition in elemental cerium although on a smaller scale.
The low-temperature $\cal{S}$-phase is suppressed by less than 500~mK at 1.9~GPa which implies that the associated QCP and the appended field-induced $\cal{L}$-to-$\cal{H}$ transition will also not be shifted greatly in this pressure range.
This, in turn, means that the absence of a strong increase in effective mass towards lower fields at 0.6 and 1.0~GPa is explained either by a change in character of the $\cal{S}$-phase boundary that makes quantum fluctuations harder to excite, or by a proximity of the $\cal{H}$-phase to a QCP in the $p$-$H$ plane of the phase diagram that is caused by fractionalization of $f$-electron spin and charge instead of symmetry breaking.
The latter scenario would be in line with the observed change of the $\cal{H}$-phase FS size with pressure in the course of a delocalisation of the 4$f$ electrons.


\section*{Acknowledgments}

We thank T.~Orton and P.~Ruddy for technical assistance.
This project has received funding from the European Research Council (ERC) under the European Union’s Horizon 2020 research and innovation programme (grant agreement No 681260) and from the EPSRC. Work performed at the at the National High Magnetic Field Laboratory, USA, is supported by NSF Cooperative Agreements DMR-1157490 and DMR-1644779, the State of Florida, U.S. DoE, and through the DoE Basic Energy Science Field Work Project Science in 100 T. Work at CSU-Fresno is supported by NSF DMR-1905636 and at UCSD by US DOE DE-FG02-04ER46105 and NSF DMR-1810310. Data presented in this article resulting from the UK effort will be made available at [URL will be added].
\vspace{0.5cm}

\bibliography{COS_pressure_Bib}

\providecommand{\noopsort}[1]{}\providecommand{\singleletter}[1]{#1}%
\begin{thebibliography}{10}

\bibitem{jeitschko_1977}
Jeitschko, W. and Braun, D.
\newblock {\em Acta Crystallographica Section B: Structural Crystallography and
  Crystal Chemistry}{ \bf 33}(11), 3401--3406 November  (1977).

\bibitem{braun_1980}
Braun, D.~J. and Jeitschko, W.
\newblock {\em Journal of Solid State Chemistry}{ \bf 32}(3), 357--363 May
  (1980).

\bibitem{Ho_2005}
Ho, P.-C., Yuhasz, W.~M., Butch, N.~P., Frederick, N.~A., Sayles, T.~A.,
  Jeffries, J.~R., Maple, M.~B., Betts, J.~B., Lacerda, A.~H., Rogl, P., and
  Giester, G.
\newblock {\em Physical Review B}{ \bf 72}(9), 094410 September  (2005).

\bibitem{bauer_2002}
Bauer, E.~D., Frederick, N.~A., Ho, P.-C., Zapf, V.~S., and Maple, M.~B.
\newblock {\em Physical Review B}{ \bf 65}(10), 100506 February  (2002).

\bibitem{bochenek_2012}
Bochenek, L., Wawryk, R., Henkie, Z., and Cichorek, T.
\newblock {\em Physical Review B}{ \bf 86}(6), 060511 August  (2012).

\bibitem{sekine_1997}
Sekine, C., Uchiumi, T., Shirotani, I., and Yagi, T.
\newblock {\em Physical Review Letters}{ \bf 79}(17), 3218--3221 October
  (1997).

\bibitem{sekine_1998}
Sekine, C., Saito, H., Uchiumi, T., Sakai, A., and Shirotani, I.
\newblock {\em Solid State Communications}{ \bf 106}(7), 441--445 May  (1998).

\bibitem{cao_2004}
Cao, D., Bridges, F., Chesler, P., Bushart, S., Bauer, E.~D., and Maple, M.~B.
\newblock {\em Physical Review B}{ \bf 70}(9), 094109 September  (2004).
\newblock Number: 9.

\bibitem{meisner_1985}
Meisner, G.~P., Torikachvili, M.~S., Yang, K.~N., Maple, M.~B., and Guertin,
  R.~P.
\newblock {\em Journal of Applied Physics}{ \bf 57}(8), 3073--3075 April
  (1985).

\bibitem{shirotani_1999}
Shirotani, I., Uchiumi, T., Sekine, C., Hori, M., Kimura, S., and Hamaya, N.
\newblock {\em Journal of Solid State Chemistry}{ \bf 142}(1), 146--151 January
   (1999).

\bibitem{grandjean_1984}
Grandjean, F., Gérard, A., Braung, D.~J., and Jeitschko, W.
\newblock {\em Journal of Physics and Chemistry of Solids}{ \bf 45}(8),
  877--886 January  (1984).

\bibitem{bauer_2001}
Bauer, E.~D., Slebarski, A., Freeman, E.~J., Sirvent, C., and Maple, M.~B.
\newblock {\em Journal of Physics: Condensed Matter}{ \bf 13}(20), 4495 (2001).

\bibitem{yogi_2005}
Yogi, M., Kotegawa, H., Zheng, G.-q., Kitaoka, Y., Ohsaki, S., Sugawara, H.,
  and Sato, H.
\newblock {\em Journal of the Physical Society of Japan}{ \bf 74}(7),
  1950--1953 July  (2005).

\bibitem{ho_2016}
Ho, P.~C., Singleton, J., Goddard, P.~A., Balakirev, F.~F., Chikara, S.,
  Yanagisawa, T., Maple, M.~B., Shrekenhamer, D.~B., Lee, X., and Thomas, A.~T.
\newblock {\em Physical Review B}{ \bf 94}(20), 205140 November  (2016).

\bibitem{gotze_unusual_2020}
G\"{o}tze, K., Pearce, M.~J., Goddard, P.~A., Jaime, M., Maple, M.~B., Sasmal,
  K., Yanagisawa, T., McCollam, A., Khouri, T., Ho, P.-C., and Singleton, J.
\newblock {\em Physical Review B}{ \bf 101}(7), 075102 February  (2020).

\bibitem{tayama_2015}
Tayama, T., Ohmachi, W., Wansawa, M., Yutani, D., Sakakibara, T., Sugawara, H.,
  and Sato, H.
\newblock {\em Journal of the Physical Society of Japan}{ \bf 84}(10), 104701
  September  (2015).

\bibitem{iwasa_magnetic_2008}
Iwasa, K., Itobe, S., Yang, C., Murakami, Y., Kohgi, M., Kuwahara, K.,
  Sugawara, H., Sato, H., Aso, N., Tayama, T., and Sakakibara, T.
\newblock {\em Journal of the Physical Society of Japan}{ \bf 77}(Suppl.A),
  318--320 January  (2008).

\bibitem{harima_2003}
Harima, H. and Takegahara, K.
\newblock {\em Journal of Physics: Condensed Matter}{ \bf 15}(28), S2081
  (2003).

\bibitem{yan_2012}
Yan, B., M\"{u}chler, L., Qi, X.-L., Zhang, S.-C., and Felser, C.
\newblock {\em Physical Review B}{ \bf 85}(16), 165125 April  (2012).

\bibitem{namiki_2003}
{Namiki}, T., {Aoki}, Y., {Sugawara}, H., and {Sato}, H.
\newblock {\em Acta Physica Polonica B}{ \bf 34}, 1161 February  (2003).

\bibitem{sugawara_2005}
Sugawara, H., Osaki, S., Kobayashi, M., Namiki, T., Saha, S.~R., Aoki, Y., and
  Sato, H.
\newblock {\em Physical Review B}{ \bf 71}(12), 125127 March  (2005).

\bibitem{rotundu_2006}
Rotundu, C.~R. and Andraka, B.
\newblock {\em Physical Review B}{ \bf 73}(14), 144429 April  (2006).

\bibitem{helm_2020}
Helm, T., Grockowiak, A.~D., Balakirev, F.~F., Singleton, J., Betts, J.~B.,
  Shirer, K.~R., König, M., Förster, T., Bauer, E.~D., Ronning, F., Tozer,
  S.~W., and Moll, P. J.~W.
\newblock {\em Nature Communications}{ \bf 11}(1), 3482 July  (2020).

\bibitem{singleton}
Singleton, J.
\newblock {\em Reports on Progress in Physics}{ \bf 63}(8), 1111--1207 jul
  (2000).

\bibitem{Thomas_2020}
Thomas, S.~M., Santos, F.~B., Christensen, M.~H., Asaba, T., Ronning, F.,
  Thompson, J.~D., Bauer, E.~D., Fernandes, R.~M., Fabbris, G., and Rosa, P.
  F.~S.
\newblock {\em Science Advances}{ \bf 6}(42), eabc8709 October  (2020).

\bibitem{goddard_2002}
Goddard, P., Tozer, S.~W., Singleton, J., Ardavan, A., Abate, A., and Kurmoo,
  M.
\newblock {\em Journal of Physics: Condensed Matter}{ \bf 14}(31), 7345--7361
  July  (2002).

\bibitem{shishido_2005}
Shishido, H., Settai, R., Harima, H., and Ōnuki, Y.
\newblock {\em Journal of the Physical Society of Japan}{ \bf 74}(4),
  1103--1106 April  (2005).

\bibitem{okada_2008}
Okada, H., Igawa, K., Takahashi, H., Kamihara, Y., Hirano, M., Hosono, H.,
  Matsubayashi, K., and Uwatoko, Y.
\newblock {\em Journal of the Physical Society of Japan}{ \bf 77}(11), 113712
  November  (2008).

\bibitem{drozdov_2015}
Drozdov, A.~P., Eremets, M.~I., Troyan, I.~A., Ksenofontov, V., and Shylin,
  S.~I.
\newblock {\em Nature}{ \bf 525}(7567), 73--76 September  (2015).

\bibitem{Clune_2020}
Clune, A., Harms, N., O’Neal, K.~R., Hughey, K., Smith, K.~A., Obeysekera,
  D., Haddock, J., Dalal, N.~S., Yang, J., Liu, Z., and Musfeldt, J.~L.
\newblock {\em Inorganic Chemistry}{ \bf 59}(14), 10083--10090 July  (2020).

\bibitem{Bangura_2007}
Bangura, A.~F., Goddard, P.~A., Singleton, J., Tozer, S.~W., Coldea, A.~I.,
  Ardavan, A., McDonald, R.~D., Blundell, S.~J., and Schlueter, J.~A.
\newblock {\em Physical Review B}{ \bf 76}(5), 052510 August  (2007).

\bibitem{Ghannadzadeh_2013}
Ghannadzadeh, S., Möller, J.~S., Goddard, P.~A., Lancaster, T., Xiao, F.,
  Blundell, S.~J., Maisuradze, A., Khasanov, R., Manson, J.~L., Tozer, S.~W.,
  Graf, D., and Schlueter, J.~A.
\newblock {\em Physical Review B}{ \bf 87}(24), 241102 June  (2013).

\bibitem{ghannadzadeh_2011}
Ghannadzadeh, S., Coak, M., Franke, I., Goddard, P.~A., Singleton, J., and
  Manson, J.~L.
\newblock {\em Review of Scientific Instruments}{ \bf 82}(11), 113902 November
  (2011).

\bibitem{sakakibara_contactless_1989}
Sakakibara, T., Goto, T., and Miura, N.
\newblock {\em Review of Scientific Instruments}{ \bf 60}(3), 444--449 March
  (1989).

\bibitem{coniglio_improvements_2010}
Coniglio, W.~A., Winter, L.~E., Rea, C., Cho, K., and Agosta, C.~C.
\newblock {\em arXiv:1003.5233 [physics]}{ \bf } March  (2010).
\newblock arXiv: 1003.5233.

\bibitem{thompson_1983}
Thompson, J.~D., Fisk, Z., Lawrence, J.~M., Smith, J.~L., and Martin, R.~M.
\newblock {\em Physical Review Letters}{ \bf 50}(14), 1081--1084 April  (1983).

\bibitem{gignoux_1985}
Gignoux, D. and Voiron, J.
\newblock {\em Physical Review B}{ \bf 32}(7), 4822--4824 October  (1985).

\bibitem{drymiotis_2005}
Drymiotis, F., Singleton, J., Harrison, N., Lashley, J.~C., Bangura, A.,
  Mielke, C.~H., Balicas, L., Fisk, Z., Migliori, A., and Smith, J.~L.
\newblock {\em Journal of Physics: Condensed Matter}{ \bf 17}(7), L77 (2005).

\bibitem{sugawara_high_2004}
Sugawara, H., Kobayashi, M., Kuramochi, E., Osaki, S., R.~Saha, S., Namiki, T.,
  Aoki, Y., and Sato, H.
\newblock {\em Journal of Magnetism and Magnetic Materials}{ \bf 272-276},
  E115--E116 May  (2004).

\bibitem{sato_transport_2003}
Sato, H., Aoki, Y., Namiki, T., Matsuda, T.~D., Abe, K., Osaki, S., Saha,
  S.~R., and Sugawara, H.
\newblock {\em Physica B: Condensed Matter}{ \bf 328}(1), 34--38 April  (2003).

\bibitem{Hall}
K. G\"otze \textit{et al}., \textit{unpublished}.

\bibitem{shoenberg}
Shoenberg, D.
\newblock {\em Magnetic Oscillations in Metals}.
\newblock Cambridge University Press, Cambridge, England,  (1984).

\bibitem{hedo_pressure_2003}
Hedo, M., Uwatoko, Y., Sugawara, H., and Sato, H.
\newblock {\em Physica B: Condensed Matter}{ \bf 329-333}, 456--457 May
  (2003).

\bibitem{rotundu_exotic_2007}
Rotundu, C.~R., Andraka, B., and Schlottmann, P.
\newblock {\em Physical Review B}{ \bf 76}(5), 054416 August  (2007).

\bibitem{ito_microscopic_2010}
Ito, T.~U., Higemoto, W., Ohishi, K., Satoh, K., Aoki, Y., Toda, S., Kikuchi,
  D., Sato, H., and Baines, C.
\newblock {\em Physical Review B}{ \bf 82}(1), 014420 July  (2010).

\bibitem{he_2011}
He, H.-T., Wang, G., Zhang, T., Sou, I.-K., Wong, G. K.~L., Wang, J.-N., Lu,
  H.-Z., Shen, S.-Q., and Zhang, F.-C.
\newblock {\em Physical Review Letters}{ \bf 106}(16), 166805 April  (2011).

\bibitem{liu_2012}
Liu, M., Zhang, J., Chang, C.-Z., Zhang, Z., Feng, X., Li, K., He, K., Wang,
  L.-l., Chen, X., Dai, X., Fang, Z., Xue, Q.-K., Ma, X., and Wang, Y.
\newblock {\em Physical Review Letters}{ \bf 108}(3), 036805 January  (2012).

\bibitem{yang_2013}
Yang, Q.~I., Dolev, M., Zhang, L., Zhao, J., Fried, A.~D., Schemm, E., Liu, M.,
  Palevski, A., Marshall, A.~F., Risbud, S.~H., and Kapitulnik, A.
\newblock {\em Physical Review B}{ \bf 88}(8), 081407 August  (2013).

\bibitem{zheng_2016}
Zheng, G., Wang, N., Yang, J., Wang, W., Du, H., Ning, W., Yang, Z., Lu, H.-Z.,
  Zhang, Y., and Tian, M.
\newblock {\em Scientific Reports}{ \bf 6}(1), 21334 February  (2016).

\bibitem{rebar_2019}
Rebar, D.~J., Birnbaum, S.~M., Singleton, J., Khan, M., Ball, J.~C., Adams,
  P.~W., Chan, J.~Y., Young, D.~P., Browne, D.~A., and DiTusa, J.~F.
\newblock {\em Physical Review B}{ \bf 99}(9), 094517 March  (2019).

\bibitem{gschneider}
Koskenmaki, D.~C. and Gschneider, K.~A.
\newblock {\em Handbook on the Physics and Chemistry of Rare Earths}, volume~1.
\newblock edited by K. A. Gschneider and L. Eyring (North Holland, Amsterdam),
  (1978).
\newblock p. 337 ff.

\bibitem{dzero_2000}
Dzero, M.~O., Gor'kov, L.~P., and Zvezdin, A.~K.
\newblock {\em Journal of Physics: Condensed Matter}{ \bf 12}(47), L711--L718
  November  (2000).

\bibitem{cecoin5}
Maksimovic, N., Eilbott, D.H., Cookmeyer, T., Wan, F., Rusz, J., Nagarajan, V.,
  Haley, S.C., Maniv, E., Gong, A., Faubel, S., Hayes, I.M., Bangura, A.,
  Singleton, J., Palmstrom, J.C., Winter, L., McDonald, R., Jang, S., Ai, P.,
  Lin, Y. Ciocys, S., Gobbo, J., Werman, Y., Oppeneer, P.M., Altman, E.,
  Lanzara, A., and Analytis, J.G. \textit{Science} (2021), \textit{in press}.

\end{thebibliography}
\bibliographystyle{nature} 


\end{document}